\documentclass{JINST}
\usepackage[utf8x]{inputenc}
\usepackage{lineno}

\newcommand{\gerda}     {{\sc Gerda}}

\title{Optical fiber read-out for liquid argon scintillation light}

\author{J. Janicsk\'o Cs\'athy$^a$\thanks{Corresponding author.}~ T.Bode$^a$, J. Kratz$^a$, S. Sch\"onert$^a$, and Ch. Wiesinger$^a$\\
\llap{$^a$}Physik Department and Excellence Cluster Universe, Technische Universit\"at M\"unchen\\ 
M\"unchen, Germany\\

  E-mail: \email{janicsko@mytum.de} 
}

\abstract{
In this paper we describe the performance of a light detector for Ar scintillation light made of wavelength-shifting (WLS) fibers 
connected to Silicon-Photomultipliers (SiPM). The setup was conceived to be used as anti-Compton veto for high purity germanium (HPGe) 
detectors operated directly in liquid Argon (LAr). Background suppression efficiencies for different radioactive sources were measured 
in a test cryostat with about 800 kg LAr. This work was part of the R\&D effort for the \gerda{} experiment. }

\begin{document}
%\maketitle

\section{Introduction}

Presently there is a growing global demand for LAr scintillation light read-out
 on the field of dark matter and high energy neutrino beam experiments. \cite{DUNE}, \cite{DarkSide},\cite{DEAP}. 
 In this paper we present a solution that uses commercial products and is suited for low background experiments.    
%Detection of the scintillation light in LAr it still a challenging task especially for low background experiments. 

Although PMT technology made a significant improvement recently the low temperature and the low radioactivity requirement 
of some experiments is still posing a problem.
A possible alternative to PMTs operated in the cryoliquid are wavelength-shifting (WLS) fibers that collect the light and guide it to a Photo-Multiplier (PMT) outside the volume. 
An example of such a setup can be found in \cite{McKinsey}. 

Alternatively one can place the photon detector in the cryo-liquid. 
For such application a low radioactivity detector is needed that works reliably at low temperatures.
The best candidate to date is the Silicon-Photomultipliers (SiPM) that is known to work at 
low temperature and because it's small size it could easily satisfy the radioactivity requirements.
The typically few mm$^2$ active surface of the SiPM naturally suggests an application with WLS fibers.

%It is known that SiPMs are working well at cryogenic temperatures but they have a rather small active surface typically a few mm$^2$ only. 
%Without further development SiPMs alone can not provide the necessary coverage for a large volume scintillator detector. 
%Therefore the straightforward solution is to combine SiPMs with WLS fiber. 

%Naturally rises the question if such a detector can compete against old fashioned PMTs. 
%In this paper we will try to demonstrate that the low detection efficiency can be compensated 
%with coverage and the resulting photo-electron (p.e.) yield  and background suppression efficiencies are comparable with 
%the results achieved in similar detectors equipped with PMTs. 

Our first attempt to build such a detector for LAr scintillation light with WLS fibers and SiPMs was described in \cite{MPI_paper}. 
The setup described in this paper is a scaled up and improved version. It is also new that the setup was realised 
without reflector foil only with WLS fiber allowing for a much larger instrumented LAr volume.

\section{Experimental setup}
%\label{sec:infrastructure}

The experiments described in this paper were conducted in the underground laboratory of the TUM in Garching (Germany). 
The laboratory building was planed to provide infrastructure for several cryogenic and low background experiments 
and it has a soil overburden of 10 m water equivalent. 

\subsection{Test stand} 
 
\begin{figure}
\centering 
\includegraphics[width=0.9\columnwidth]{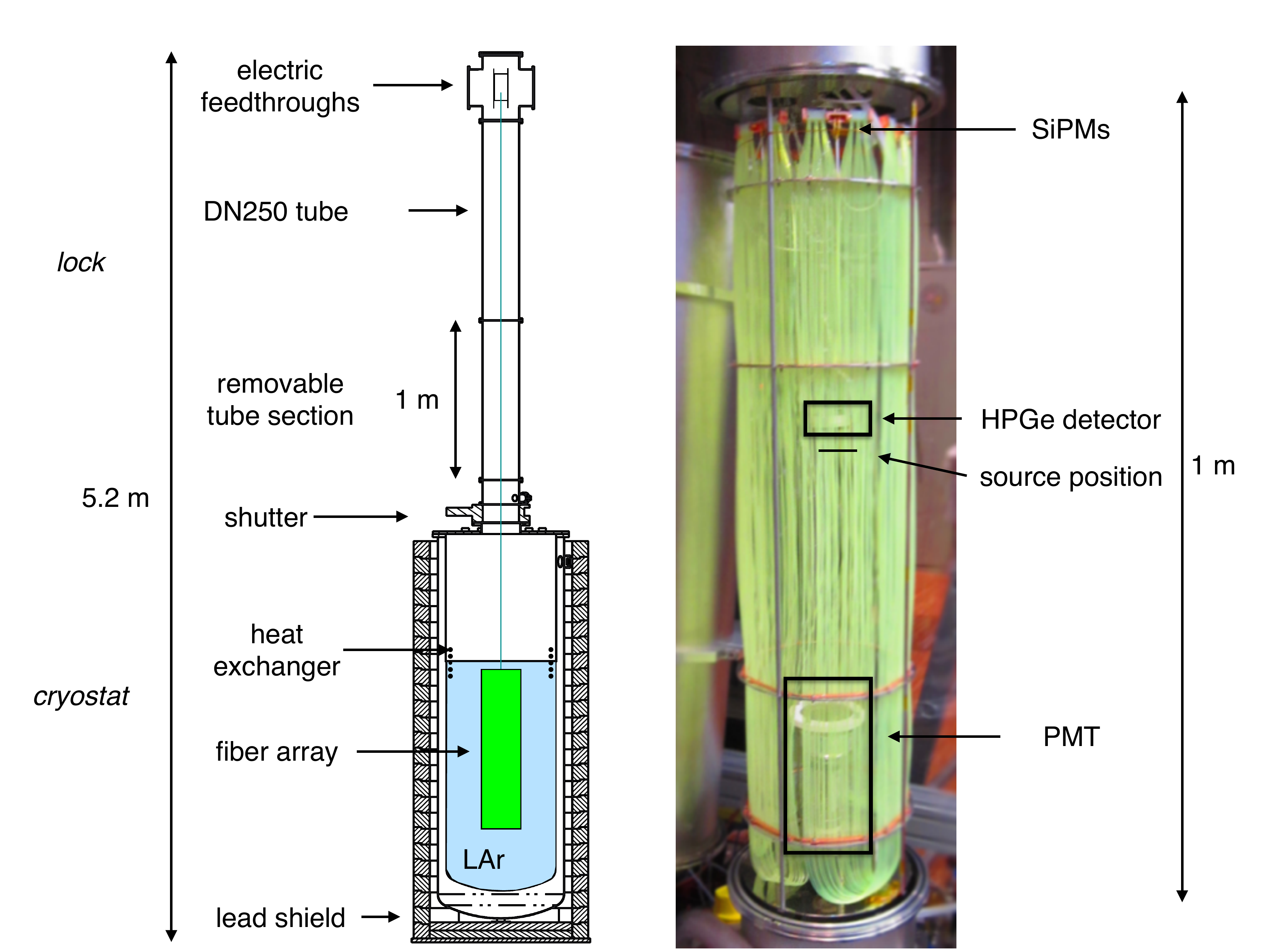}
\caption{Schematic drawing of the cryostat (left) and the fiber array mounted in the lock before lowering in the cryostat (right).}
\label{fig:larsetup}
\end{figure}

In the underground laboratory is a test cryostat with a capacity of about 600 l of LAr. 
The cryostat has an internal diameter of 700 mm and the liquid level is at about 1.5 m from the bottom
allowing the submersion of 1 m high objects. The schematic drawing of the cryostat is shown in Fig.\ref{fig:larsetup}.   

The LAr is actively cooled with liquid nitrogen. 
During the experiments described here and previous tests we achieved 
stable operation of several month without the need to refill LAr.

The cryostat is equipped with an air tight lock similar to the one used in the \gerda{} experiment \cite{gerda}. 
The lock permits lowering different setups in the LAr without contaminating it with air.  
The largest object that can be inserted through the lock has a diameter of 240 mm and a height of one meter.

The experimental setup described in Sec.\ref{sec:lightinst} is suspended with a steel band that is wind up 
on a pulley actuated by an electric motor. Electrical connections are provided via vacuum feed troughs located 
on the top of the lock. The signal transmission lines are realised with 50 $\Omega$ coaxial cables. 
  
The steel band and the cables are long enough to reach the bottom of the cryostat even without the detectors mounted. 
The total length of the signal cables from the detectors to the electronics rack is about 20 m. 

\subsection{Liquid Argon}

The LAr was delivered by the company Air Liquide and it was certified to be better than N50 quality. 

\begin{figure}
\centering \includegraphics[width=0.5\columnwidth]{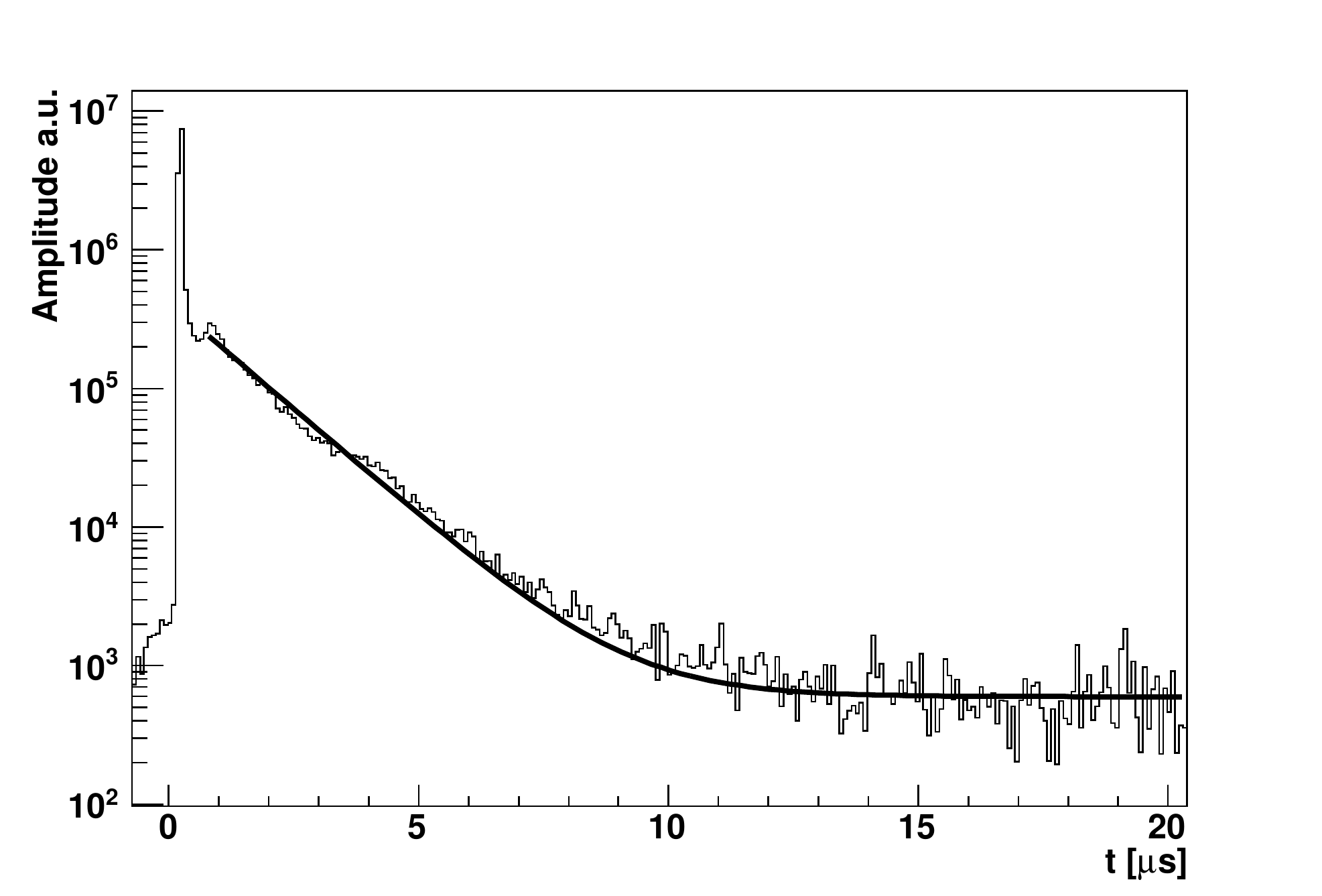}
\caption{Sum of PMT waveforms recorded in LAr. The solid line indicates the fitted exponential plus a constant (flat background). 
The decay time of the exponential curve is about 1.4 $\mu$s.}
\label{fig:triplet}
\end{figure}

The impurities in the LAr have an effect on the scintillation light yield 
affecting in first place the light emitted from the triplet state of the Ar excimer. 
Therefore the lifetime of the triplet state is a good measure of the purity of the Argon. 
Fig.\ref{fig:triplet} shows summed up PMT waveforms recorded during our experiments.
The fitted exponential indicates a the triplet life time of around 1.4 $\mu$s which is exceptionally good for such an experiment 
without active purification of the Argon. The literature value of the triplet life time is 1.6 $\mu$s \cite{Hitachi}  

\subsection{Germanium detector}

For the experiments described below we needed a high-purity germanium (HPGe) detector that we could operate directly in the LAr. 
The used detector was of BEGe (Broad Energy Germanium) detector type from Canberra with only 50 g mass.
%and previously was used for electronic tests.

The  HPGe detector was mounted in a self-made holder and it was suspended to hang at about the middle of the setup described 
in Sec.\ref{sec:array}. Both the high-voltage and signal contacts were realised with wedge bonding. 
The signal contact was connected to a cryogenic amplifier similar to the one used in \gerda{}  \cite{gerda}.
During our experiments the energy resolution was about 3 keV FWHM at the 2.6 MeV line of $^{208}$Tl. 

\subsection{Radioactive sources}

%1. low activity sources to avoid pile up of the events.
%2. sources for cryoliquid
%3. open alpha/beta sources
All low background experiments suffer from the presence of radioactive impurities 
 in the surrounding materials usually dominated by isotopes from the Th and U decay chain. 
Therefore we tested the response of our detector to $^{232}$Th and $^{226}$Ra sources.  
The purpose of the described experiments was to demonstrate the background reduction potential of a similar setup in the \gerda{} experiment.
 
Since most of the radioactive sources at our disposition were not certified for use in cryo-liquids 
we used thorated welding rods as a $^{232}$Th source. 
Commercially available tungsten welding rods contain natural $^{232}$Th oxide up to 8 \%. 
The measured activity of the 1 mm , 10 cm long welding rod was about 20 Bq. 

To simulate uranium chain contamination we used $^{226}$Ra source with about 30 Bq activity. 
This source was a bare metal plate with Ra deposited on the surface, an open alpha and beta emitter. 

The sources were suspended at about 1 cm below the HPGe detector.

\section{Light instrumentation}
\label{sec:lightinst}

The light detector described in this paper was designed with regard to the very specific requirements of the \gerda{} experiment. 
These requirements are: low activity materials, low weight and sufficiently large instrumented volume for an 
effective anti-Compton veto. Another design constraint was that the light instrumentation should be inserted in the 
cryostat together with the germanium detectors hence the diameter is limited by the neck of the  \gerda{} cryostat.  

The major difference compared to the setup described in \cite{MPI_paper} that we omitted the reflector foil.
Instead we tried to achieve a closed volume using only WLS fibers. 
With the fibers arranged on the perimeter of a cylinder it is possible to detect light also 
from outside of the cylinder increasing the volume of the LAr compared to a setup delimited by reflector material. 
This way we can compensate for the geometrical limitations of our lock 
which does not allow for an instrumented volume with a radius larger than the radiation length of the LAr (14 cm).
In \gerda{} the situation is similar, there the distance from the HPGe detectors to the fibers would be sub 
optimal for an anti-Compton veto. This arrangement implies that the LAr volume is larger than the instrumented volume.

Below we describe the components used in our setup and the way it was built. 

\subsection{WLS fibers}

Optical fibers are usually made with round cross section. It is known that an optical fiber with square cross section
has a larger acceptance angle because of simple geometrical reasons. The acceptance angle of the fiber
which is proportional to the trapping efficiency for scintillating and WLS fibers is
also affected by the cladding of the fiber. Fibers with two cladding layers with decreasing refraction index from inside
to outside have higher trapping efficiency than the more common single clad fibers.

For the highest possible trapping efficiency we chose to use square, multiclad fibers from Saint-Gobain.

The scintillation light of LAr (128 nm) is difficult to detect directly therefore the VUV light  
is shifted to blue (~430 nm) by Tetraphenyl-butadiene (TPB). The TPB is applied as a coating on the fiber.

\begin{figure}
\centering \includegraphics[width=0.5\columnwidth]{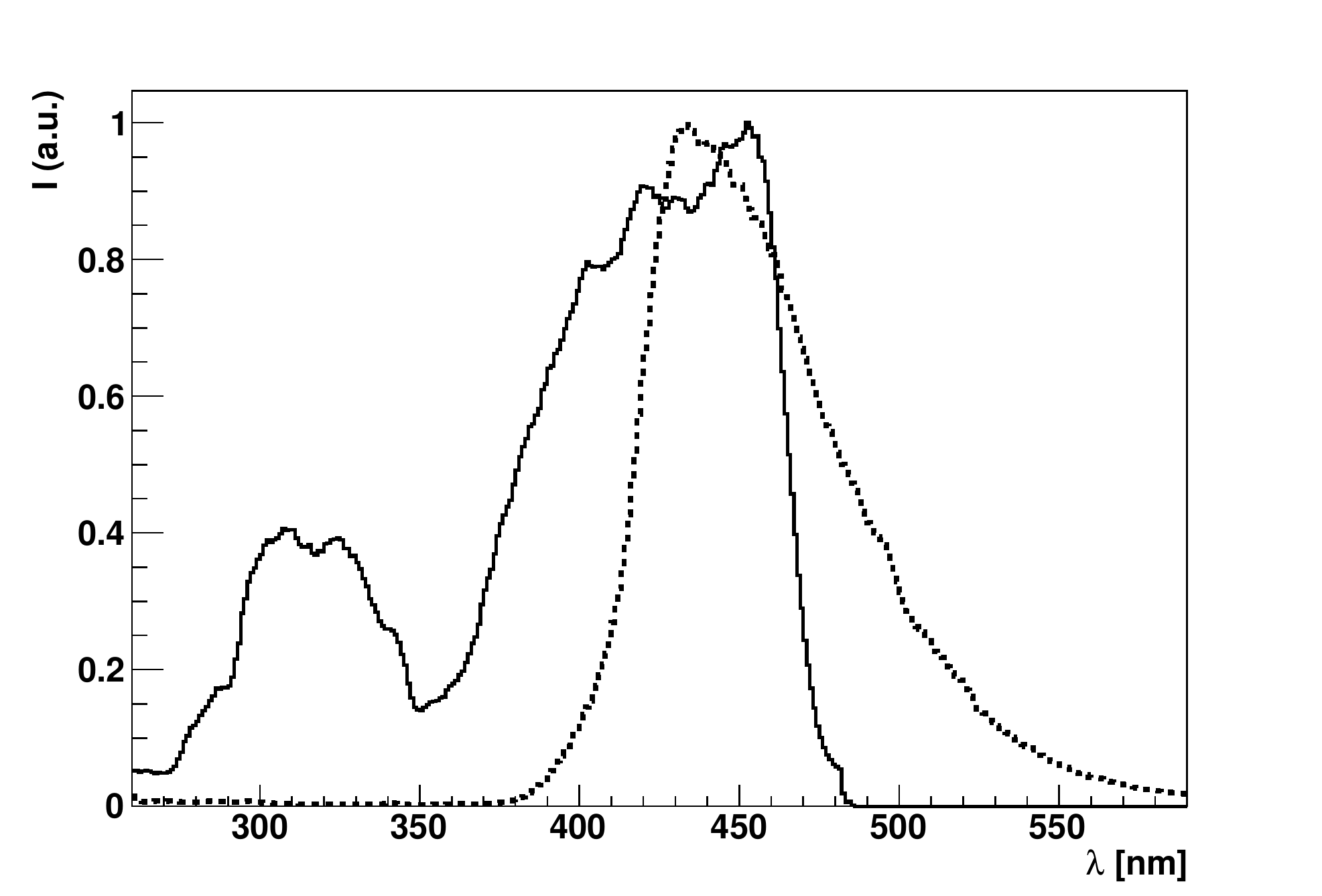}
\caption{Absorption spectrum of the BCF-91A fiber (solid line) and emission spectrum of the TPB (dashed line).}
\label{fig:tpbSpec}
\end{figure}

The emission spectrum of the TPB was measured together with the absorption spectrum of the WLS fiber
with a Carry Eclipse fluorimeter. From the two type of WLS fibers offered by Saint-Gobain it was found
that the BCF-91A fiber matches better the emission spectrum of TPB (see Fig.\ref{fig:tpbSpec}). The graphical
comparison of the two spectra shows that about 60\% of the light emitted by the TPB can be
absorbed by the fiber.

The purchased fibers were characterised for their optical properties. 
With the help of an LED pulser and a SiPM we measured the attenuation length of the fiber.  
The measurement resulted 3.8 m which is consistent with the $>$3.5 m given by the manufacturer. 
To estimate the efficiency of the fiber capturing blue light 
a piece of fiber was dissolved in toluene and we measured the extinction coefficient of the solution 
with a Perkin-Elmer UV-VIS spectrometer. The measured result corresponds to an absorption length of the fiber of 0.7$\pm$0.001 mm 
for blue (400 nm) light. 
 %only 76% is absorbed ?

\subsection{Fiber modules}
\label{sec:modules}

Since nine 1x1 mm$^2$ fibers cover perfectly the surface of a 3x3 mm$^2$ SiPM the number fibers in the array should be multiple of nine.
For easier handling we chose a modular design with units of 54 fibers.  
 
\begin{figure}
%\centering \includegraphics[width=0.5\columnwidth]{pix/copper_holder.png}
%\includegraphics[width=0.5\columnwidth]{pix/holder.jpeg}
\includegraphics[width=\columnwidth]{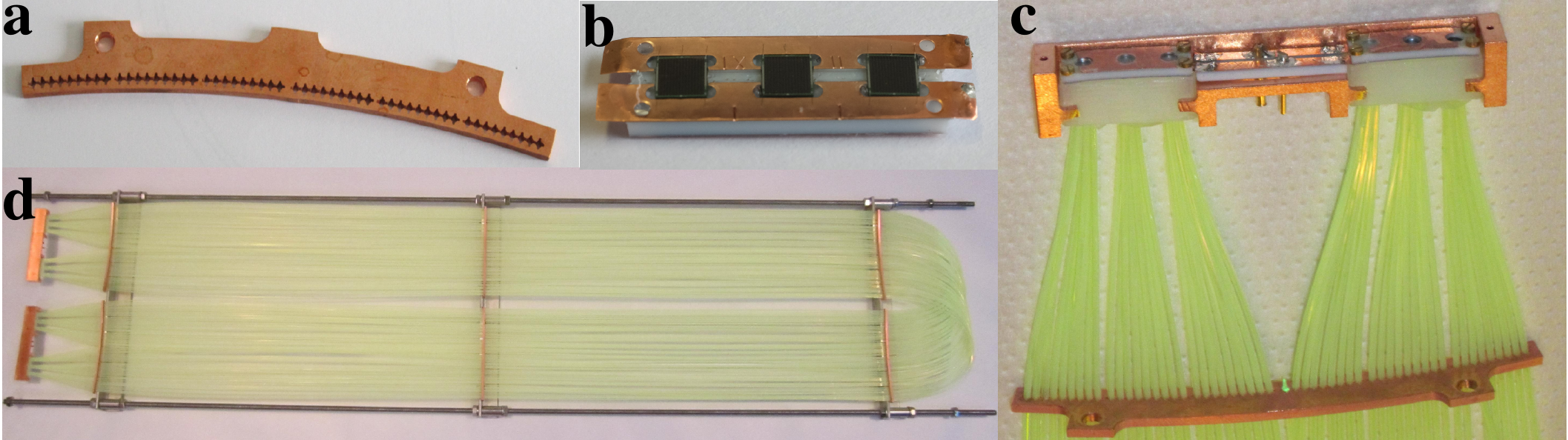}
\caption{a) copper holder that keeps the fibers turned at 45$^\circ$, b) self made SiPM array of 3 SiPMs, c) SiPM coupling to WLS fibers, d) completed double fiber module on a mounting frame. Note the fibers from one module bent into the other.}
\label{fig:copper_holder}
\end{figure}

To achieve maximal coverage with minimum quantity of WLS fiber the square fibers are arranged such that their diagonal 
lies on the circumference of the cylinder we want to enclose. 
Hence the visible fiber surface facing the inner volume is 1.41 mm times the length of the fiber.  
To prevent them turning, the fibers are fed through a holder shown in Fig.\ref{fig:copper_holder}. 
This holder also fixes the fiber modules to the supporting structure.

At the lower end of the fiber module the fibers are not cut but they are bent 180$^o$ to return in the adjacent module. 
This way we don't need light detectors or mirrors at the lower end of the setup. 
All SiPMs are on the top and both ends of the fibers are read-out. 

The length of a fiber module is one meter. Because the fibers turn back in the next module the total length of a single fiber is
slightly less then two meters. The losses because of the attenuation in the fiber have the effect that we collect about 10\% 
less light than if we would place SiPMs also at the lower end of the module.   

%2 m length 76.2 \% of the light collected assuming uniform illumination
%1m length 87\% ....
%10\% less light as if we would instrument both ends of the fiber of 1 m length

\subsection{TPB coating}

One possibility to apply TPB coating on the fibers is to dip them in a solution of TPB and polystyrene dissolved in some
organic solvent. This method was used in previous experiments to coat mirror foil \cite{LArGe_paper, MPI_paper}.
%and also was applied to fibers \cite{MPI_paper}. 
The solvent of choice is usually toluene.
It was noticed that the solvent damaged the fibers causing cracks in the cladding where the fiber was bent.

After some moderately successful experiments we abandoned the dip coating in favor of vacuum deposition. 
Vacuum deposition of TPB is done regularly on reflector materials and we adapted the technology to WLS fibers.

\begin{figure}
%\centering \includegraphics[width=0.30\columnwidth]{pix/evap_lowres.png}\includegraphics[width=0.5\columnwidth]{pix/evap_in_lowres.png}
\centering \includegraphics[height=8cm]{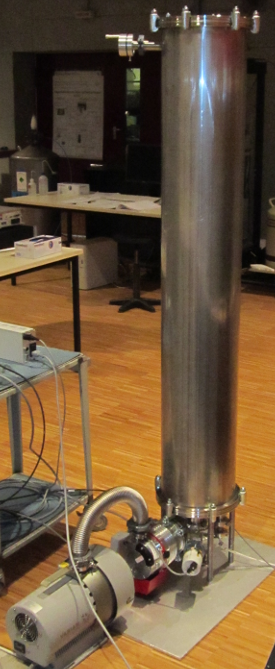}\includegraphics[height=8cm]{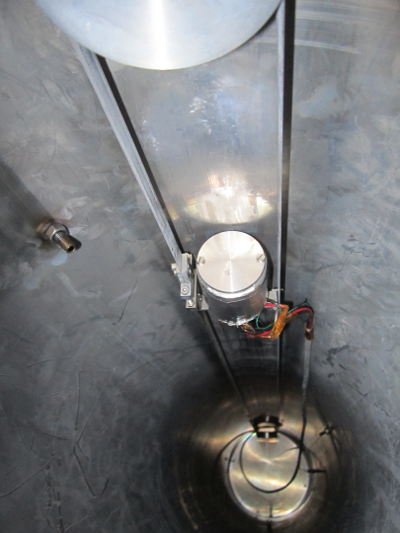}
\caption{Left: TPB evaporator. Right: heated crucible inside the vacuum chamber.Fixed on the steel band the crucible can be moved up and down by turning the pulley.}
\label{fig:tpbEvap}
\end{figure}

The fiber modules were coated with TPB in a vacuum chamber consisting of a 250 mm diameter tube (Fig.\ref{fig:tpbEvap}, left). 
Inside the tube we installed a heated crucible made of aluminium. The crucible was closed on the top and had a slit machined 
on the side such that the TPB vapors could exit only in the direction of the target. 
The picture in Fig.\ref{fig:tpbEvap} shows the crucible inside the vacuum chamber mounted on a steel band.
The steel band forms a closed loop stretched between two pulleys.
By turning one of the wheels the crucible moves down and covers the full length of the fiber module with TPB.

The most uniform deposition we achieved by heating the crucible to temperatures well below the melting point
of the TPB (203.5 $^o$C). At 170 $^o$C the deposition was reasonably fast and the resulting TPB layer had 
satisfactory mechanical and optical properties. 

%\begin{figure}
%\centering \includegraphics[width=0.8\columnwidth]{pix/tpb_thick_2.pdf}
%\label{fig:tpbThick}
%\caption{Scintillation light yield and sample transparency versus the evaporation time at 175 $^o$C.
%Blue squares: transmission, red dots: fluorescence intensity at 430 nm}
%\end{figure}

The thickness of the deposited TPB was measured by evaporating it on glass plates. 
The plates with TPB layer were scratched and we measured the depth of the scratch with a Tencor P-10 Surface Profiler.

For optical characterisation we evaporated TPB on
acrylic plates exposing them for a different time at the same temperature.
The relative fluorescence intensity of the exposed plates was measured with a Carry Eclipse fluorimeter.
The optical transmission of the same plates was measured with a Perkin-Elmer UV-VIS spectrometer.

%The thickness of the TPB layer evaporated on the fibers was chosen to empirically such that the 
%transmission of the coated plate was still about 80\%. 

Based on the  results of these measurement we chose to evaporate about 200 nm TPB on the fibers. 
At this thickness the scintillation light yield is already very high and the transmission 
of the coating is still about 80\%. Thicker TPB layers where white and opaque. 

%The results are shown in Fig.\ref{fig:tpbThick}.
%One can see that the light yield is not increasing after an exposure of 20 min. while the 
%transparency of the samples is decreasing continuously. 
%One can see that the light yield is increasing with the deposition time while the
%transparency of the samples is decreasing continuously.
%Measure light yield in transmission mode ? 

During the evaporation of the fiber modules the TPB thickness was monitored with glass plates exposed together with
the fibers. The TPB thickness on the final fiber modules was varying between 200 and 400 nm.

%The measured TPB thicknes is documented for each fibers module installed in GERDA. See Table in Appendix 1.

%\begin{figure}
%\centering \includegraphics[width=0.5\columnwidth]{pix/fibre_module_evap.jpeg}
%\label{fig:fibmodule}
%\caption{Fibre module as taken out from the evaporator. The glass plates used for thickness monitoring are fixed in the middle.The
%measured TPB thickness for each module is given in Appendix 1. }
%\end{figure}
   
%\begin{figure}
%\centering \includegraphics[width=0.5\columnwidth]{pix/coupling.jpeg}
%\label{fig:coupling}
%\caption{54 WLS fibers coupled to 6 SiPMs.}
%\end{figure}                             

\subsection{SiPM read out}
 
SiPMs were extensively studied at low temperatures: \cite{Otono,Lightfoot:2008im,Akiba,MPI_paper} 
The result of this studies was always that SiPMs are a viable option at temperatures down to that of liquid nitrogen (LN). 
A big advantage of using SiPMs at LN temperatures is that the dark rate decreases by 6-7 orders 
of magnitudes to an almost negligible level. 
This fact allows us to consider SiPM arrays with an active area of the order of cm$^2$ for scintillation light read out in LAr. 

For possible use in \gerda{} the radioactivity of commercial devices is a major concern because
of the substrate that is normal glass fiber circuit board material.
To have the radio-purity issue under control we decided to package the SiPMs by ourselves.
Therefore we purchased the SiPMs in die from Ketek GmbH. We used the 3 mm x 3 mm devices 
with 50 $\mu$m pixel size from Ketek with product code PM3350.

Due to the temperature dependence of the ploy-silicon quenching resistor the recharge 
of the pixel at low temperature is slower typically by an order of magnitude.
The quenching resistor of these devices increased from 360 k$\Omega$ to about 4 M$\Omega$ 
causing that the time needed to recharge a pixel increased to about two $\mu$s. 

The slow pulses do not have any drawback since the goal of the experiment is to detect 
a single photo-electron in a time window of 5 to 10 $\mu$s. 

%Although the pulse shapes are different all tested devices behave well at LN temperature.

%A disadvantage of SiPMs is the small sensitive area which we tried to overcome by attaching them to WLS fibers.
%The Hamamatsu 3mm x 3mm comes on a PCB like board with epoxy glue on the top.
%The price is high, about 200 Euro/piece. Buying large quantities (100 piece) would not reduce the price (Situation 2013).

%Another reason for trying a self made packaging is the thick glue on the top of the commercial SiPMs.
%Because the light exits the fiber under an angle of 27.4$^o$ (entering a medium with a refraction index of 1.6)
%the 0.3 mm epoxy layer on the top of the SiPM will cause a light loss of up to 18\%.
%(9, 1 mm$^2$ fiber coupled to 3x3 mm$^2$ SiPM).  With the self made packaging we
%set a design goal of 0.1 mm gap between the SiPM sensitive surface and the polished end of the fiber.
%With a 0.1 mm gap the light loss could be as low as 6.5\%. Unfortunately this design goal was compromised by the deformation
%of the (Teflon based) Cuflon material used as substrate.

As substrate for the packaging CuFlon, a Teflon based circuit board material was chosen (
CuFlon is a brand name of the Polyflon Company). This material is known for its low activity.  

In a small board of 7 mm x 21 mm we had square holes machined for the SiPM chips.
Such a holder with the SiPMs already implanted is shown in Fig.\ref{fig:copper_holder}.b).
Each holder has place for three SiPMs. The top copper layer of the PCB material is divided
in two strips during the milling to form the two contacts of the SiPM array.

The SiPMs are placed in the holder and bonded to the copper stripes. Than the holder is covered with
a thin layer of transparent epoxy glue (Polytec EP601).
The SiPM arrays are placed in an oven heated to 40 $^o$C for about 10 hours, until the glue is cured.

Each array was tested first at room temperature than in LN. The arrays that passed the first test were assembled in double arrays of
six SiPMs and tested again in LN. Only fully functional SiPM arrays with low dark rate were assembled with the fibers.

%In a separate test it was measured if the packaging made at TUM has any effect on the p.d.e. of the devices.
%In Fig. \ref{fig:pde_package} is shown the p.d.e. of a chip packaged at TUM and on the right a the same measurement
%done with the commercial packaging. The conclusion is that our method has no negative effects on the p.d.e.
%The measurement was done at Ketek GmbH.

%Radioactivity is not a major concern since silicon should be very clean.

%\begin{figure}
%\centering \includegraphics[width=0.5\columnwidth]{pix/SiPM_package.png}
%\caption{Cuflon holder with SiPMs}
%\label{fig:cuflonHolder}
%\end{figure}

%\begin{figure}
%\centering \includegraphics[width=0.9\columnwidth]{pix/fib_module.jpeg}
%\caption{Fiber module on the mounting frame.}
%\label{fig:fibModule}
%\end{figure}

\subsection{Detector array}
\label{sec:array}
The setup shown in Fig.\ref{fig:larsetup} (right) is a scaled down version of the setup built for \gerda{}. 
The diameter of the cylindrical fiber array is 240 mm, about half of the 470 mm planed/built for \gerda{} 
while the height is exactly the same, one meter.
  
The setup comprises seven fiber modules described in Sec.\ref{sec:modules}
%Each fiber module with 54 fibers is 
%connected to an array of six 3x3 mm$^2$ SiPMs which form one read-out channel (See Fig.\ref{fig:coupling}). 
In the middle of cylinder we installed the HPGe detector. The radioactive source was suspended just below the HPGe detector.  

At the bottom part of the cylinder we mounted a PMT which was used only to measure 
the triplet life time of the LAr shown in Fig.\ref{fig:triplet}.

\subsection{Electronics and data acquisition}

\begin{figure}
\centering \includegraphics[width=0.8\columnwidth]{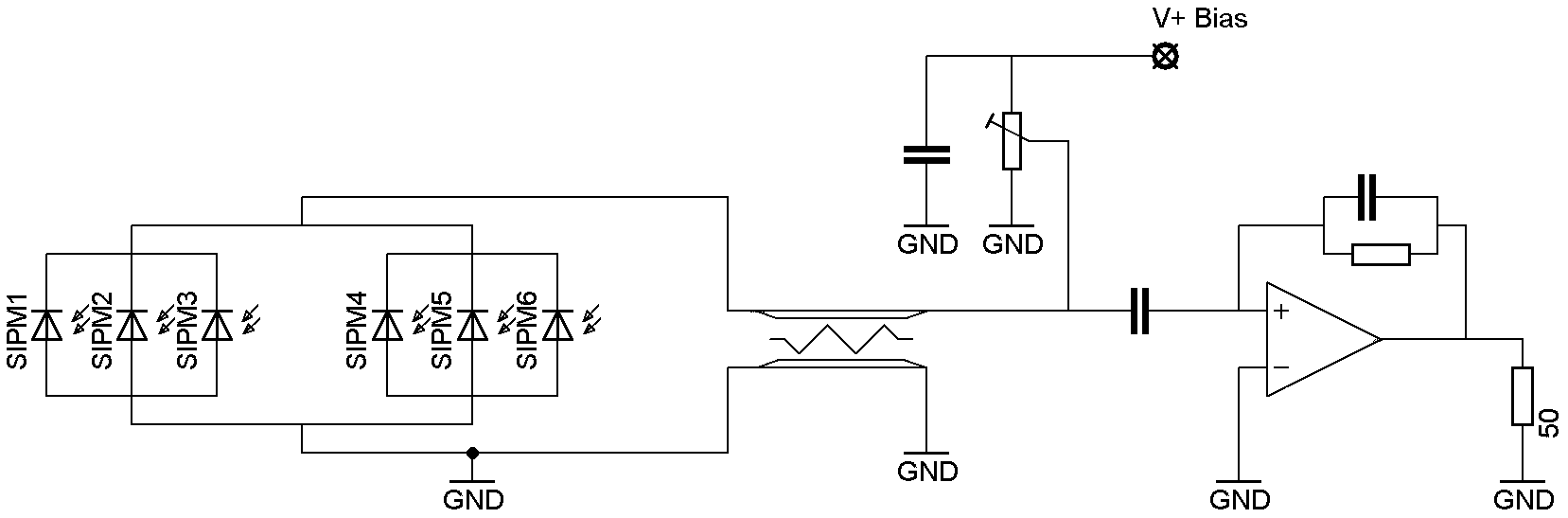}
\caption{Circuit diagram of the SiPM read-out. Only one channel is shown with six SiPMs in parallel which corresponds to an array of 54 mm$^2$. 
The cable separating the SiPMs and the amplifier is about 20 m long. The charge amplifier is a Cremat-112.}
\label{fig:sipm_readout}
\end{figure}

The HPGe detector was connected to a cryogenic amplifier similar to the one described in \cite{gerda}. 
On the contrary SiPMs arrays were operated without active electronic components in the LAr cryostat. 
The 6 SiPMs in parallel are connected directly to a 50 $\Omega$ coaxial cable which is used for applying 
the bias voltage and for signal transmission. The amplifiers were outside the cryostat at room temperature. 
A schematic drawing of the SiPM read-out is shown in Fig.\ref{fig:sipm_readout}.

The HPGe signals and the seven SiPM channels, each corresponding to a fiber module, 
were recorded separately with an eight channel Struck SIS3301 FADC. The data acquisition (DAQ) 
was usually triggered by the HPGe channel and 40 $\mu$s traces were recorded with 100 MHz sampling rate. 
To estimate the light yield (see Fig.\ref{fig:lightYield}) in a separate measurement we triggered on any of the SiPM channels.
To record the data presented in Sec. \ref{sec:BiPo} the sampling rate of the ADC was reduced to 25 MHz and the 
trace length increased to the maximum allowed by the ADC.  

\subsection{Data analysis}
\label{sec:ana}
The DAQ recorded the pulse shapes of each channel and the LAr veto was applied off-line.
The HPGe energy was calculated on-line by the Struck ADC firmware. The shaping time was set to 10 $\mu$s.

\begin{figure}
\centering \includegraphics[width=0.5\columnwidth]{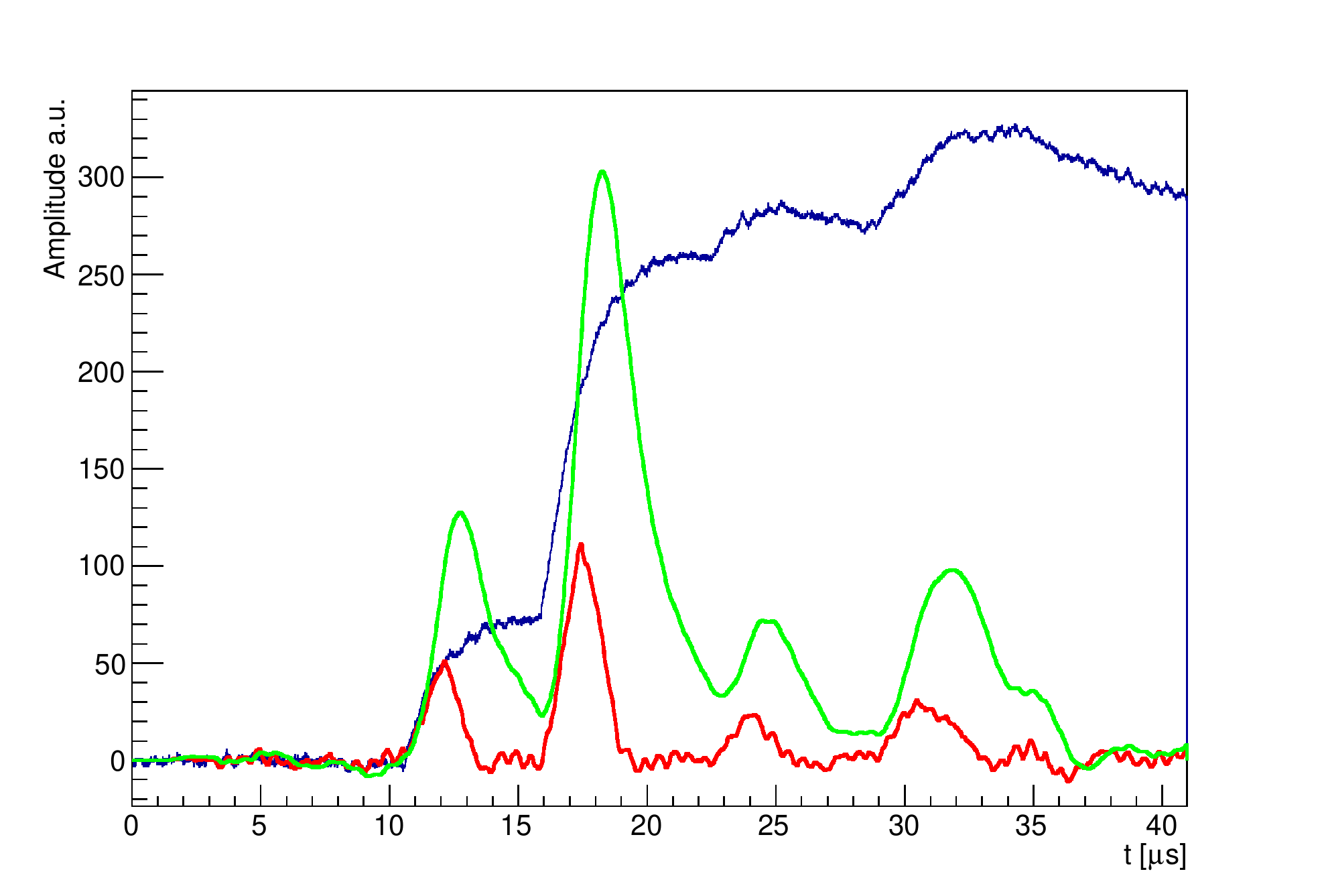}\includegraphics[width=0.5\columnwidth]{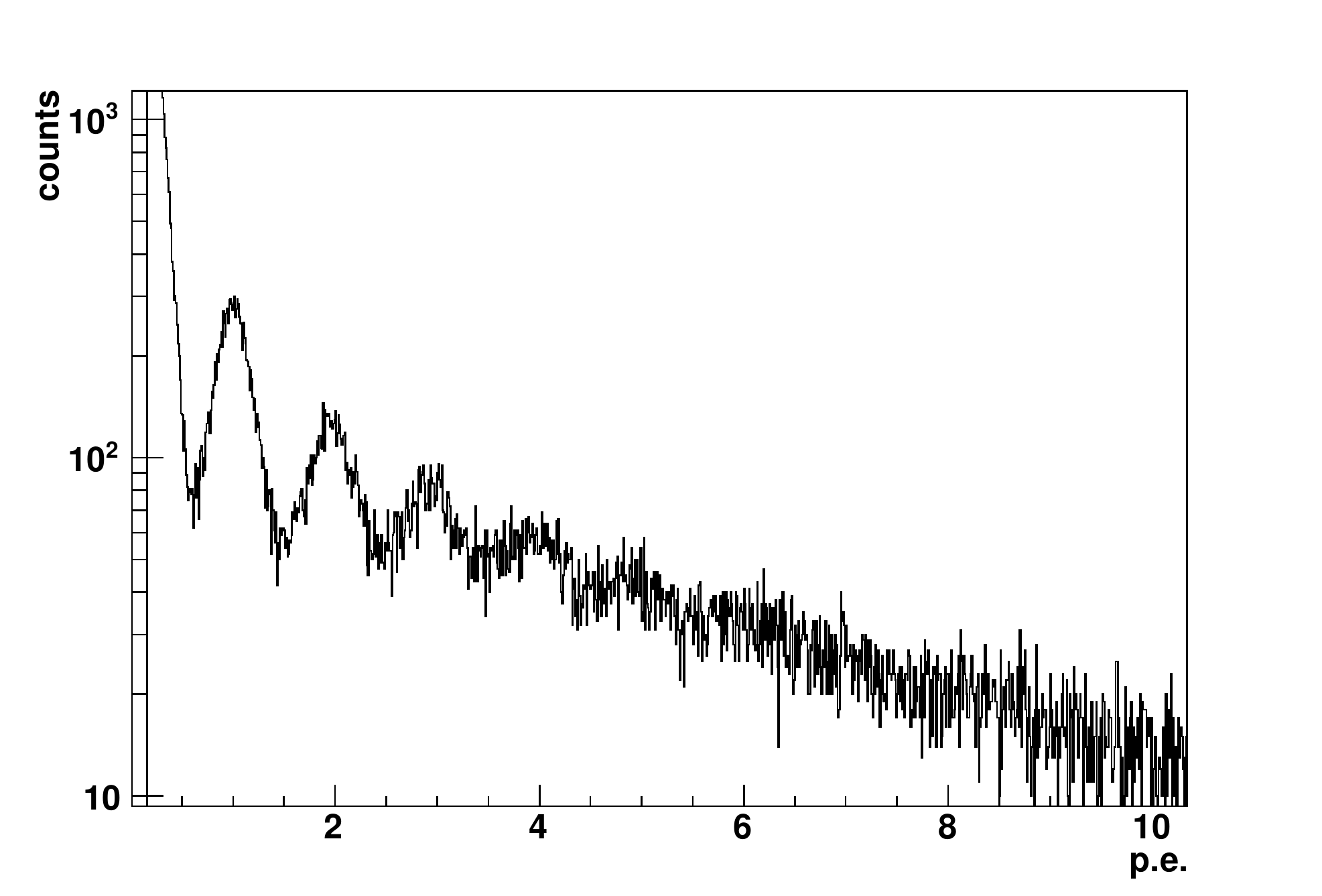}
\caption{Left: trigger filter with two moving window deconvolution. Right: Amplitude distribution of the signals after the second moving window deconvolution from one channel with 6, 3x3 mm$^2$ SiPMs. The data was recorded in LAr with $^{232}$Th source.}
\label{fig:pe_spec}
\end{figure}

To find the SiPM pulses in the recorded traces we used a trigger finding algorithm based on the well known trapezoidal filter  \cite{Stein1996141}.
We applied the mowing window deconvolution two times on the recorded traces. 
In the first step the decay time of the amplifier is deconvoluted (50 $\mu$s) and 
in the second step the RC constant of the SiPM given by the quenching resistor times pixel capacity is 
removed. On the resulting waveform we used a leading edge trigger finding algorithm to determine 
the trigger time and we read the amplitude of the pulses after a fixed delay following the trigger.    
Fig. \ref{fig:pe_spec}, left shows the original waveform and the derived waveforms and on the right the resulting amplitude distribution.

The algorithm was implemented in the GELATIO framework \cite{GELATIO} which is used to process the \gerda{} data.

For the anti-Compton veto an event was vetoed if any of the SiPM channels was hit (at least one trigger found) and the amplitude 
of the SiPM signal was at least 70\% of the single pixel signal amplitude. 
In all our experiments we used a coincidence window of 7 $\mu$s which is the time needed 
to collect 99\% of the LAr scintillation light with a triplet lifetime of 1.4 $\mu$s.
   
To produce the plot in Fig. \ref{fig:lightYield} we summed up the amplitude of coincident SiPM pulses measured with the trapezoidal filter.

\section{Experimental results}

\subsection{Photo-electron yield}

To stay consistent with the literature we will give the light yield in number of \textit{photo-electrons} (p.e.) 
although since we are using SiPMs we always mean the \textit{number of SiPM pixels fired}. 

First we lowered the setup in LAr with a built in $^{232}$Th source. 
In this measurement the DAQ was triggering on any of the SiPM channels. 

\begin{figure}
\centering \includegraphics[width=0.5\columnwidth]{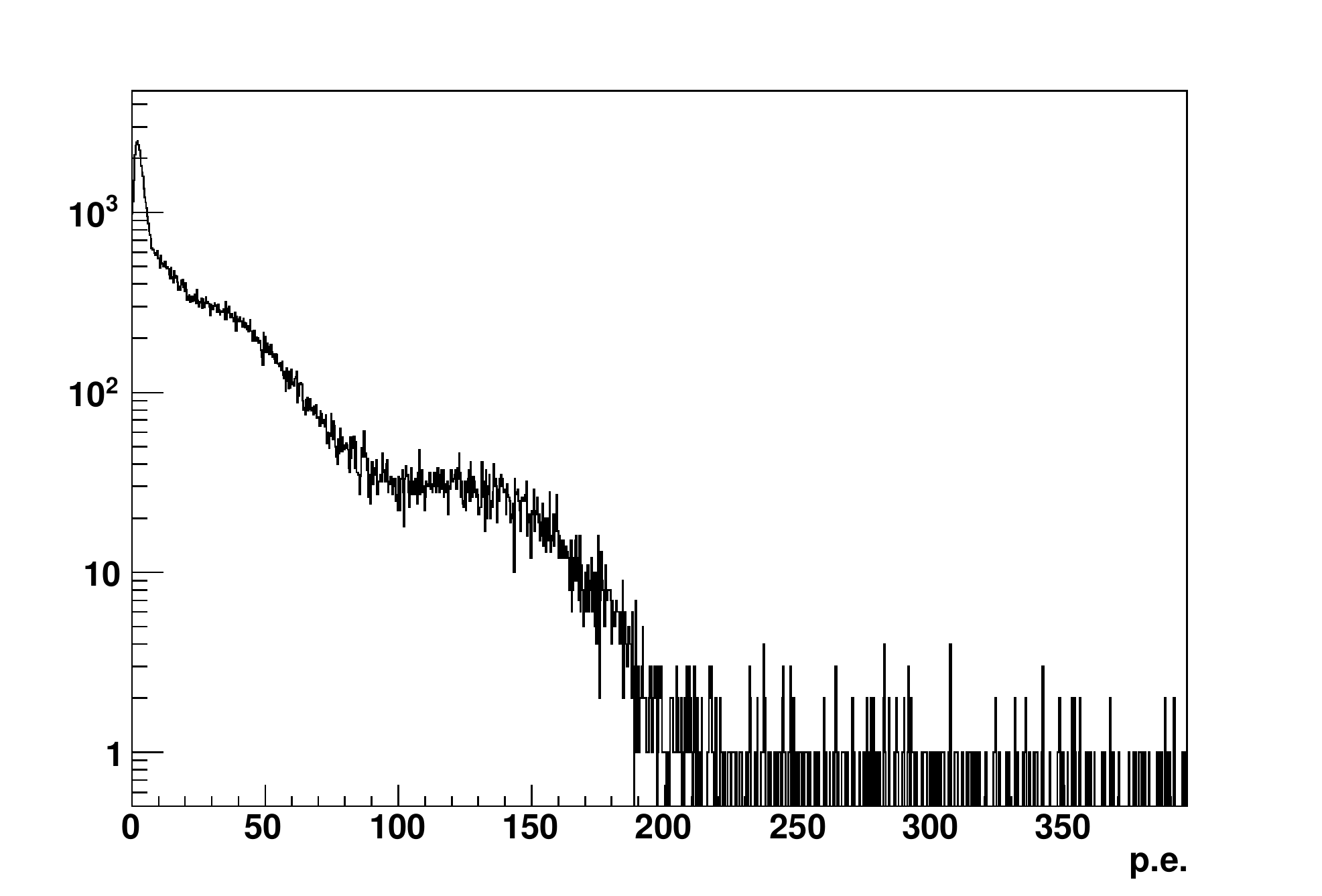} 
\caption{$^{232}$Th spectrum recorded with the fibers and SiPMs. The estimated light yield is about 50 p.e./MeV}
\label{fig:lightYield}
\end{figure}

Each channel was calibrated in p.e. than the channel amplitudes belonging to one event were summed up. 
The resulting spectrum can be seen in Fig.\ref{fig:lightYield}. 
Knowing that the highest energy line in the Th spectrum is the 2.6 MeV gamma line from the decay of $^{208}$Tl, 
one can estimate the light yield. In this case we obtained about 50 p.e./MeV. 
%(This result is valid with the 1.4 $\mu$s triplet.)

%The measured light yield is lower than what we expected (?). The difference can be explained by the lower geometrical coverage with fibers.  

\subsection{Suppression factors}

%To test the performance of our setup as anti-Compton veto we mounted a HPGe detector in the center of the fiber cylinder.  
%The DAQ channel where the HPGe detector was connected was providing the trigger and the pulse shapes of 8 channels was recorded. 
%The anti-Compton veto was applied off-line. 

To test the performance of the fibers as anti-Compton veto we mounted a $^{232}$Th source right below the HPGe detector 
before lowering the whole setup in the LAr.   

%In one experiment the fibers and the HPGe detector were submerged in the LAr together with a $^{232}$Th source. 
%The energy spectrum shown in Fig.\ref{fig:ThSuppressed} was produced  

\begin{figure}
\centering 
\includegraphics[width=0.5\columnwidth]{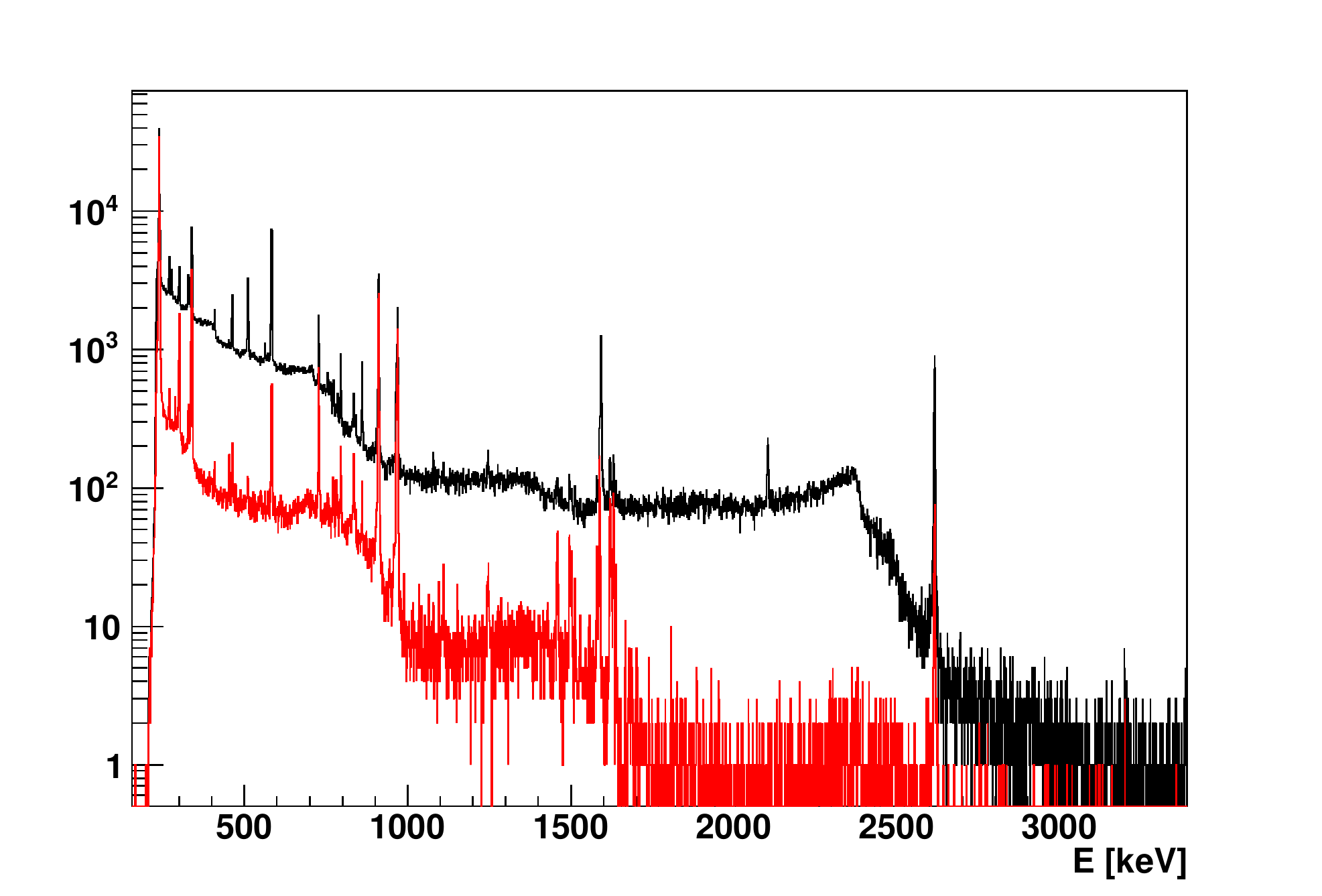}\includegraphics[width=0.5\columnwidth]{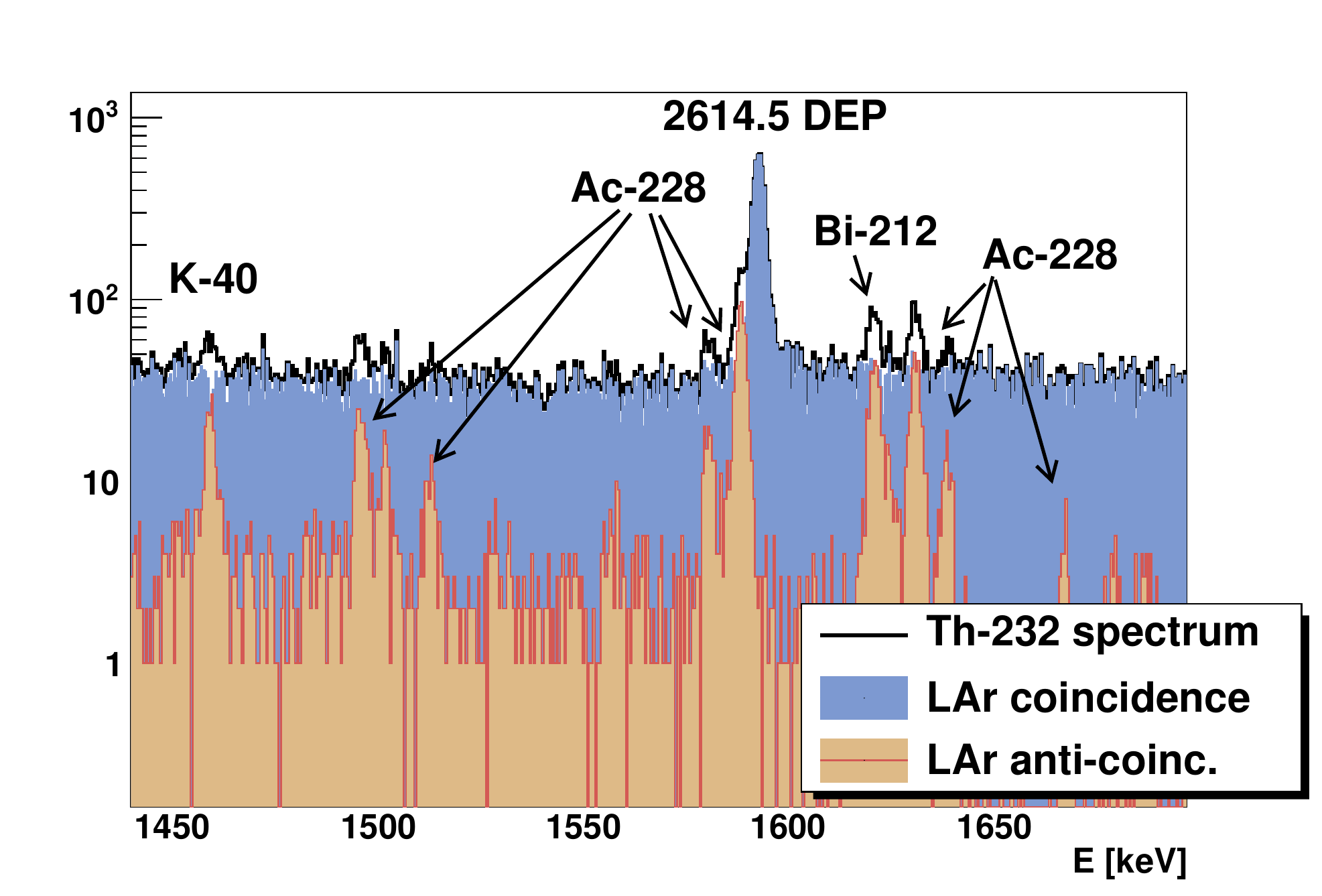}
\caption{Left:$^{232}$Th spectrum recorded with a HPGe detector and the LAr veto suppressed spectrum (red histogram). 
Right: zoom on the DEP with LAr coincidence and anti-coincidence cut. }
\label{fig:ThSuppressed}
\end{figure}

The resulting spectra can be seen in Fig.\ref{fig:ThSuppressed}. 
After applying the LAr veto cut the spectrum was suppressed by a factor 177 in window of 200 keV around 2039 keV.
In the right histogram of Fig.\ref{fig:ThSuppressed} on can see how the LAr veto brings to evidence the gamma lines
hidden under the Compton-continuum of the 2.6 MeV line of $^{208}$Tl. 
The blue filled histogram shows the spectrum of coincident events. 
The coincident spectrum above 2 MeV is contains practically only the 2.6 MeV line, 
it's Compton-continuum, the single-escape and the double escape (DEP) peaks. 
The separation of the DEP from the $^{228}$Ac line below is spectacular.  

In another experiment we submerged the setup in the LAr together with a $^{226}$Ra  source.  
The $^{226}$Ra source was an open alpha source by construction. 
An unknown fraction of the emitted alpha and beta particles could exit in the LAr.  
 
Applying the LAr veto as before we achieved a suppression of the $^{226}$Ra spectrum by a factor of about 12 
in the region of interest. The obtained suppression factors from both experiments are sumarized in Table.\ref{tab:SF}

In both experiments we estimated the probability of random coincidences between the light read-out 
and the HPGe signal by triggering the DAQ with a pulser signal and measuring the suppression of this trigger. 
In all our experiments we had a pulser acceptance rate above 90\%. 
The rate of random coincidences can be explained by the activity of our sources 
plus the activity of $^{39}$Ar in the LAr and the cosmic muon rate.

%by cutting out the gamma lines the suppression was about 15. 

\subsubsection{BiPo-214 delayed coincidences}
\label{sec:BiPo}

Scintillator detectors offer the possibility to use Bi-Po delayed coincidence for background identification. % (literature ?)

Having an open alpha source we can try to observe the alpha decay of $^{214}$Po following a hit in the 
HPGe detector induced by a $^{214}$Bi gamma photon.
 
Therefore we changed the DAQ settings to record 665 $\mu$s long traces with a sampling rate of 25 MHz.
We recorded 20 $\mu$s baseline before the HPGe trigger and the 7 $\mu$s after the trigger was used for the usual anti-Compton veto.
The remaining trace length of 628 $\mu$s (3.8 $^{214}$Po half-lives) was used to search for alpha events.  

We identified alpha events by requiring high multiplicity.
We classified a group of triggers as an alpha event when several SiPM channels fired (at least 4 of 7 ) 
and the total number of p.e.s in that event was above 20. This cut was optimised empirically in order 
to reduce the contamination of the sample with gamma background events. 
%(should be 40).

\begin{figure}
\centering 
\includegraphics[width=0.5\columnwidth]{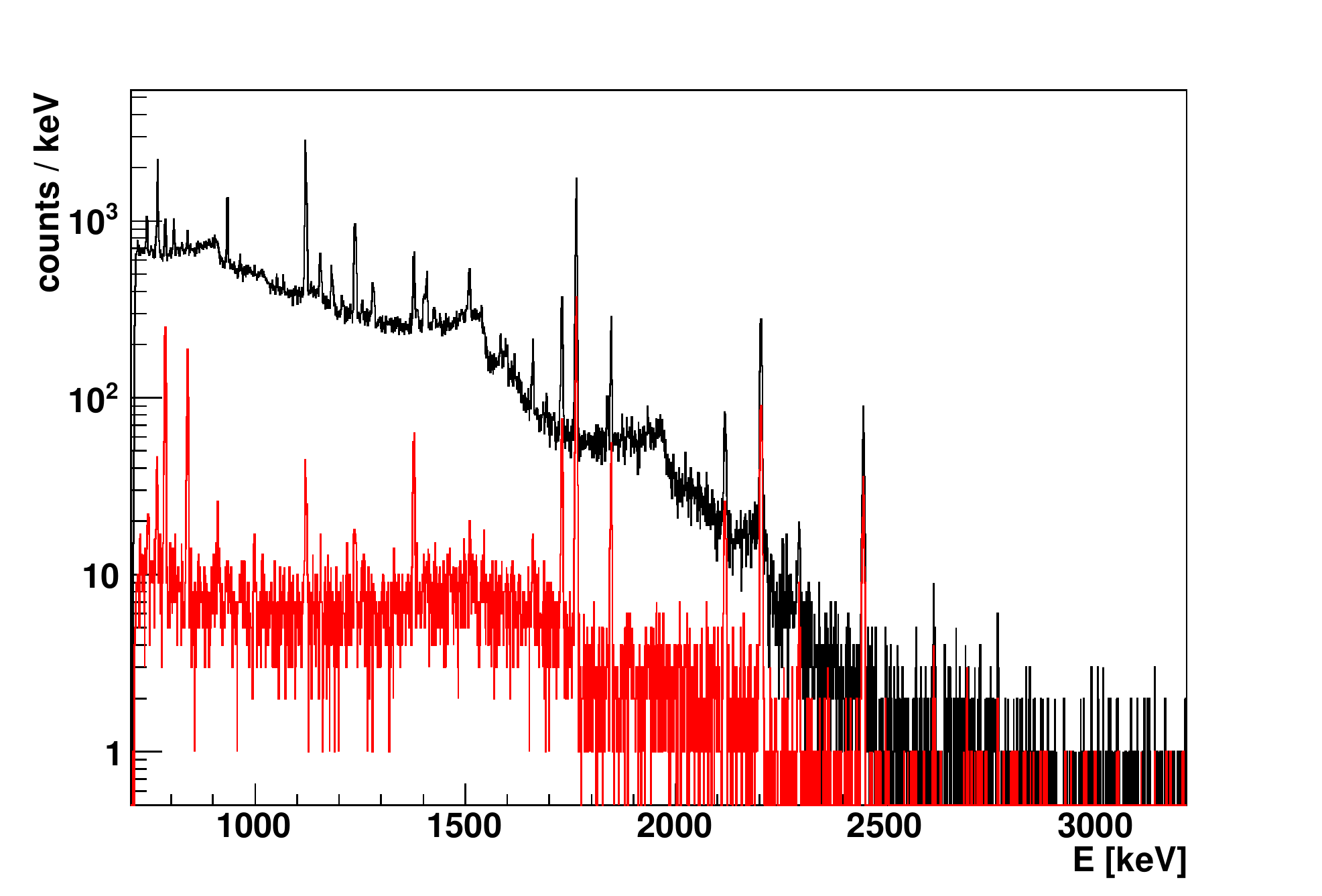}\includegraphics[width=0.5\columnwidth]{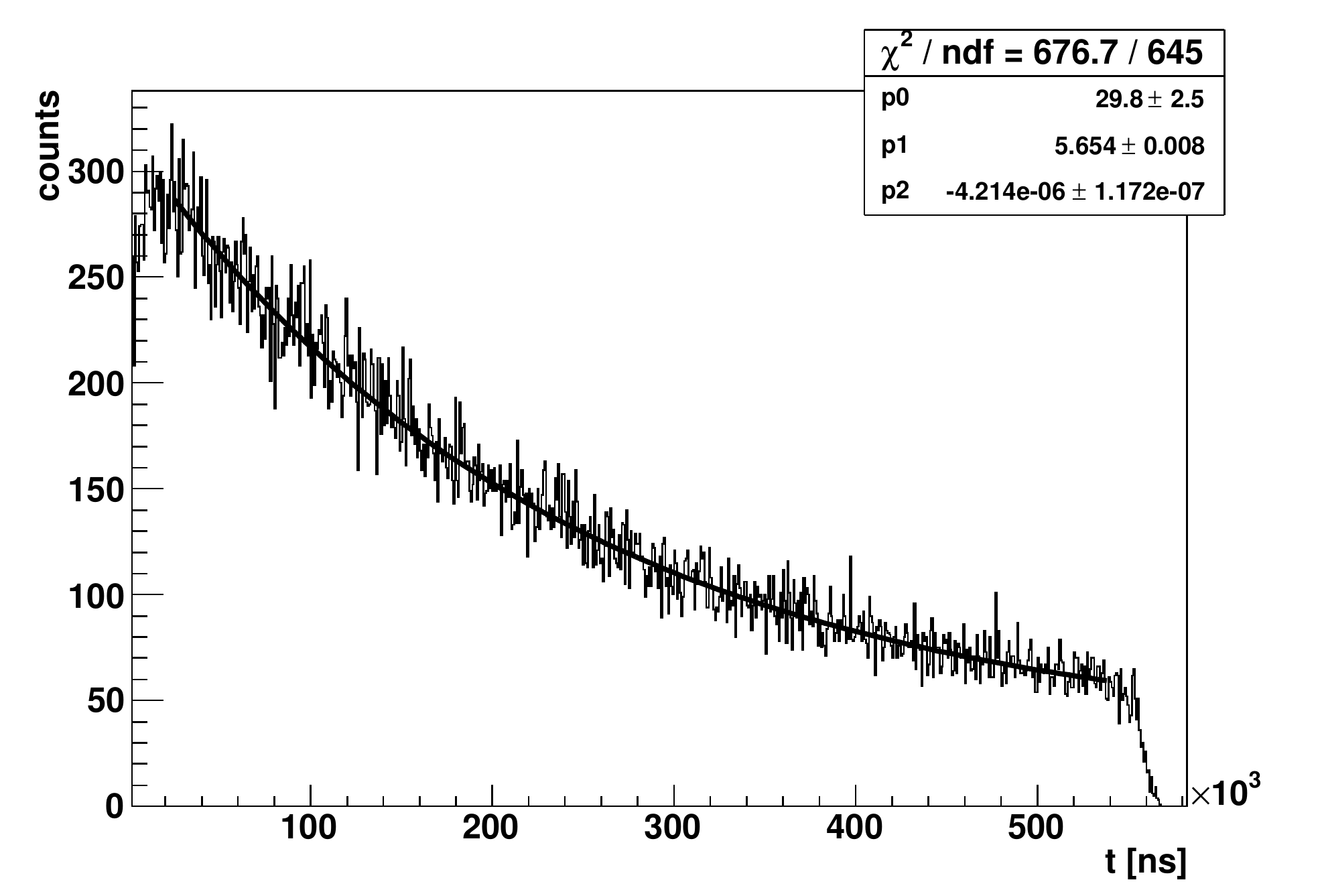}
\caption{Left: $^{226}$Ra spectrum and the suppressed spectrum. Right: time distribution of alpha events following a hit in the HPGe detector. Fitted half life: 164.5 $\pm$ 4.5  $\mu$s}
\label{fig:PoHalfLife}
\end{figure}

The number of such events is decreasing exponentially with a time constant that corresponds to a half life of 164 $\mu$s 
(See Fig.\ref{fig:PoHalfLife}). 
%Literature value 164.3 $\mu$s.
Therefore we are convinced that with the above described method we selected the alpha events originating from $^{214}$Po 
following the decay of $^{214}$Bi. 

The observation of the alpha can be also used to suppress the $^{226}$Ra spectrum in the HPGe detector. 
The measured suppression factor is only 1.3. 
We have to mention that we did not expected much better suppression: with the given source geometry the alpha 
particles are emitted in the LAr only in 50\% of the cases which sets an upper limit to the expected suppression 
factor of 2. In addition we don not know how many alphas actually exit the source without loosing too much energy.
Therefore the achieved suppression factor is reasonable and is still compatible with the assumption 
that we can detect alphas with very high probability. 

In the\gerda{} experiment the BiPo-214 trigger could be useful to reject background originating from $^{222}$Rn 
that could be either dissolved in the LAr or sticking to the surface of detectors and surrounding material. 

\begin{figure}
\centering 
\includegraphics[width=\columnwidth]{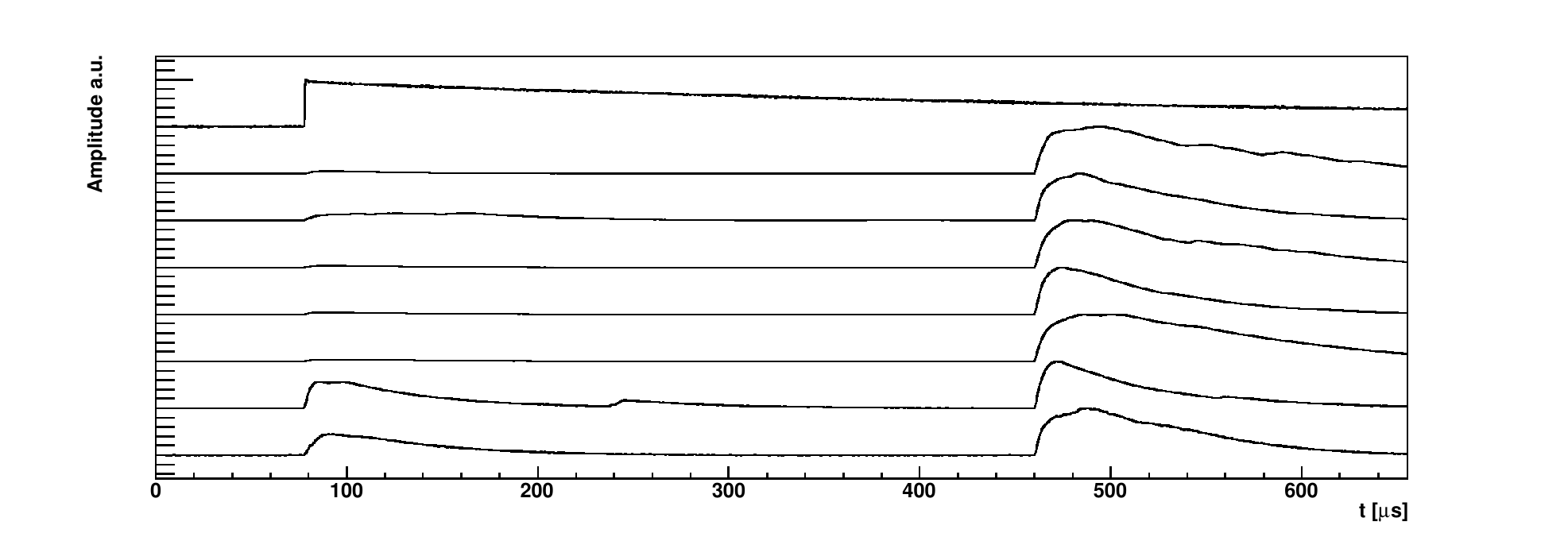}
\caption{Event display showing the stacked traces of eight channels. The trace on the top is the HPGe detector and the seven SiPM channels are below. 
One can see a prompt scintillation event in coincidence with the HPGe detector and a high energy event in the LAr about 350 $\mu$s later.}
\label{fig:bipotrig}
\end{figure}

\begin{table}[htp]
 \centering
\begin{tabular}{|c|c|c|c|}
\hline
	             & $^{232}$Th      & $^{226}$Ra & $^{226}$Ra + BiPo\\
\hline
SF 2039 $\pm$ 35 keV & 177$\pm$33 & 11.8$\pm$1.2 &  14.6$\pm$ 1.7\\
\hline
SF 2039 $\pm$ 100 keV& 150$\pm$15 & 12.2$\pm$0.7 & 15.0$\pm$0.9 \\
\hline
Pulser acceptance           & 93.7\%     & 95.2\%      &    91.3\%      \\
\hline
\end{tabular}
\caption{Measured suppression factors for$^{232}$Th and $^{226}$Ra sources.}
\label{tab:SF}
\end{table}
%No gamma line subtraction

\section{Conclusion}
In this paper we showed that is possible to build an anti-Compton veto for HPGe detectors operated in LAr using 
only WLS fibers and SiPMs. We demonstrated that the performance of such a system is very close to the one 
built with PMTs \cite{LArGe_paper}. The differences in the measured suppression factors compared to \cite{LArGe_paper}
are mainly due to the different sources used in the experiment since the p.e. yield was similar. 

The $^{228}$Th source used in \cite{LArGe_paper} had probably a higher beta emission probability. 
On the other hand our $^{226}$Ra source emitted more betas than the one used in \cite{LArGe_paper}. 
Apart from these differences the suppression factors reported in this paper are
compatible with the best results achieved elsewhere. 

Another novelty in our setup compared to similar systems is the vacuum deposition of TPB on WLS fibers. 
We obtained a coating quality that is superior to the liquid solvent based coatings and does not damage the WLS fibers.      

We also demonstrated that is possible to detect $^{214}$Bi - $^{214}$Po delayed coincidence 
with a $^{214}$Bi gamma photon triggering the HPGe detector and the following $^{214}$Po alpha particle depositing energy in the LAr. 
This could potentially provide further background reduction in experiments like \gerda{}.

%\section{Acknowledgments}
\acknowledgments

This work was supported by the Excellence Cluster and the BMBF,
The fiber holders were designed and produced at the MPIK Heidelberg under the supervision of Karl Tasso Kn\"opfle.
We are grateful for Ketek GmbH for the cooperation and for many free samples of SiPMs.  

%\begin{figure}
%\centering 
%\includegraphics[height=12cm]{pix/Cryostat2.jpeg} \includegraphics[height=12cm]{pix/250_setup.png}
%\includegraphics[width=0.15\columnwidth]{pix/cryostat.png}\includegraphics[width=0.15\columnwidth]{pix/cryo_fibers.png}
%\caption{Cryostat and fibers mounted in the lock}
%\end{figure}

%\begin{thebibliography}{1}
%\bibitem{MPI}J. Janicsk\'o Cs\'athy et al. NIM A 654 1 (2011) 225
%\bibitem{peiffer} P. Peiffer, PhD thesis (2007)
%\bibitem{gerd} G.Heuser Low-Radioactivity Background Techniques, Annu.Rev.Nucl.Part.Sci. 1995 45 543-590
%\bibitem{GEL} GELATIO reference
%\end{thebibliography}

%\bibliographystyle{model1-num-names}
%\bibliographystyle{plain}
\bibliographystyle{abbrv}
\bibliography{article}

\begin{thebibliography}{10}

\bibitem{DarkSide}
C.~E. Aalseth and {\it et al.}
\newblock The darkside multiton detector for the direct dark matter search.
\newblock {\em Advances in High Energy Physics}, 2015(541362), 2015.

\bibitem{LArGe_paper}
M.~Agostini, M.~Barnab\'e-Heider, D.~Budj\'as, C.~Cattadori, A.~Gangapshev,
  K.~Gusev, M.~Heisel, M.~Junker, A.~Klimenko, A.~Lubashevskiy, K.~Pelczar,
  S.~Schönert, A.~Smolnikov, and G.~Zuzel.
\newblock {LArGe:} active background suppression using argon scintillation for
  the {Gerda} 0$\ensuremath{\nu \beta \beta}$ -experiment.
\newblock {\em Eur. Phys. J. C}, 75(10), 2015.

\bibitem{GELATIO}
M.~Agostini, L.~Pandola, P.~Zavarise, and O.~Volynets.
\newblock {GELATIO}: a general framework for modular digital analysis of
  high-purity ge detector signals.
\newblock {\em Journal of Instrumentation}, 6(08):P08013, 2011.

\bibitem{Akiba}
M.~Akiba and {\it et al.}
\newblock Multipixel silicon avalanche photodiode with ultralow dark count rate
  at liquid nitrogen temperature.
\newblock {\em Optics Express}, 17:16885--16897, 2009.

\bibitem{DEAP}
P.-A. Amaudruz and {\it et al.}
\newblock Deap-3600 dark matter search.
\newblock {\em arXiv:1410.7673v2}.

\bibitem{McKinsey}
{D. N. McKinsey {\it et al.}}
\newblock Detecting ionizing radiation in liquid helium using wavelength
  shifting light collection.
\newblock {\em Nucl. Instr. and Meth. A}, 516(2-3):475 -- 485, 2004.

\bibitem{Hitachi}
A.~Hitachi and T.~Takahashi.
\newblock Effect of ionization density on the time dependence of luminescence
  from liquid argon and xenon.
\newblock {\em Phys. Rev. B}, 27:5279--5285, 1983.

\bibitem{MPI_paper}
{J. Janicsk\'o Cs\'athy}, {H. Aghaei Khozani}, A.~Caldwell, X.~Liu, and
  B.~Majorovits.
\newblock Development of an anti-compton veto for {HPGe} detectors operated in
  liquid argon using silicon photo-multipliers.
\newblock {\em Nuc. Instr. and Meth. A}, 654(1):225 -- 232, 2011.

\bibitem{gerda}
{K.-H. Ackermann \it et al.}
\newblock The {GERDA} experiment for the search of $0\nu\beta\beta$ decay in
  $^{76}${Ge}.
\newblock {\em Eur. Phys. J. C}, 73(3):1--29, 2013.

\bibitem{Lightfoot:2008im}
P.~K. Lightfoot, G.~J. Barker, K.~Mavrokoridis, Y.~A. Ramachers, and N.~J.~C.
  Spooner.
\newblock {Characterisation of a silicon photomultiplier device for
  applications in liquid argon based neutrino physics and dark matter
  searches}.
\newblock {\em JINST}, 3:P10001, 2008.

\bibitem{Otono}
H.~Otono and {\it et al.}
\newblock Study of {MPPC} at liquid nitrogen temperature.
\newblock {\em Proceedings of International Workshop on New Photon-Detectors
  PD07}, PD07:007, 2007.

\bibitem{DUNE}
R.Acciarri and {\it et al.}
\newblock Long-baseline neutrino facility ({LBNF}) and deep underground
  neutrino experiment ({DUNE}) conceptual design report volume 2.
\newblock {\em arXiv:1512.06148}.

\bibitem{Stein1996141}
J.~Stein, F.~Scheuer, W.~Gast, and A.~Georgiev.
\newblock X-ray detectors with digitized preamplifiers.
\newblock {\em Nuc. Instr. and Meth. B}, 113:141 -- 145, 1996.

\end{thebibliography}

\end{document}